\documentclass[a4paper,11pt]{article}
\usepackage{pos}

\newcommand{\bea}{\begin{eqnarray}} \newcommand{\eea}{\end{eqnarray}}

\def\Comment#1{}

\newcommand{\bean}{\begin{eqnarray*}}
\newcommand{\eean}{\end{eqnarray*}}

\newcommand{\gapproxeq}{\lower
.7ex\hbox{$\;\stackrel{\textstyle >}{\sim}\;$}}
\newcommand{\lapproxeq}{\lower
.7ex\hbox{$\;\stackrel{\textstyle <}{\sim}\;$}}

\newcommand\lsim{\mathrel{\rlap{\lower4pt\hbox{\hskip1pt$\sim$}}
    \raise1pt\hbox{$<$}}}
\newcommand\gsim{\mathrel{\rlap{\lower4pt\hbox{\hskip1pt$\sim$}}
    \raise1pt\hbox{$>$}}}
\newcommand{\ba}{\begin{array}}
\newcommand{\ea}{\end{array}}

\newcommand{\be}{\begin{equation}}
\newcommand{\ee}{\end{equation}}
\newcommand{\bear}{\begin{eqnarray}}
\newcommand{\eear}{\end{eqnarray}}

\newcommand{\ket}{\,\rangle}
\newcommand{\bra}{\langle \,}

\newcommand{\mY}{\mathcal{Y}}
\newcommand{\Frac}[2]{\frac{\displaystyle #1}{\displaystyle #2}}

\title{Searching for heavy resonances via oblique parameters in non-linear effective frameworks}

\author*[a,\dag]{Ignasi Rosell}
\notes{\note{We wish to thank the organizers for the pleasant conference. 
This work has been supported in part by the Spanish Government (PID2019-108655GB-I00, PID2020-114473GB-I00, PID2022-137003NB-I00, PID2023-146220NB-I00); by the Generalitat Valenciana (PROMETEU/2021/071); by the Universidad Cardenal Herrera-CEU (INDI24/17 and GIR24/16); by the ESI International Chair@CEU-UCH; by EU COMETA COST Action CA22130; by the Universidad Complutense de Madrid (research group 910309); and by the IPARCOS institute.
}}
\author[b]{Antonio Pich}
\author[c]{Juan Jos\'e Sanz-Cillero}

\affiliation[a]{Departamento de Matem\'aticas, F\'\i sica y Ciencias Tecnol\' ogicas, Universidad Cardenal Herrera-CEU, CEU Universities, 46115 Alfara del Patriarca, Val\`encia, Spain}
\affiliation[b]{IFIC, Universitat de Val\`encia -- CSIC, Apt. Correus 22085, 46071 Val\`encia, Spain}
\affiliation[c]{Departamento de F\'\i sica Te\'orica and Instituto de F\'\i sica  de  Part\'\i culas  y  del  Cosmos IPARCOS,  Universidad Complutense de Madrid, E-28040 Madrid, Spain}

\emailAdd{rosell@uchceu.es}
\emailAdd{pich@ific.uv.es}
\emailAdd{jjsanzcillero@ucm.es}

\abstract{Within the framework of a general non-linear effective field theory describing the electroweak symmetry breaking, we perform a detailed analysis of the next-to-leading contributions to the electroweak oblique parameters $S$ and $T$ from hypothetical heavy resonance states strongly coupled to Standard Model fields. This work extends our previous results by including parity-odd operators in the effective Lagrangian, contributions from fermionic cuts, and up-to-date experimental constraints. We demonstrate that in strongly-coupled ultraviolet completions satisfying both Weinberg Sum Rules —as is the case in asymptotically free gauge theories— the vector and axial-vector resonance masses are constrained to lie above $10\,$TeV. Conversely, scenarios allowing for lighter resonances with masses between $2\,$and $10\,$TeV necessarily imply a violation of the second Weinberg Sum Rule.}

\FullConference{The European Physical Society Conference on High Energy Physics (EPS-HEP2025)\\
7-11 July 2025\\
Marseille, France\\}

\begin{document}
\maketitle

\section{Introduction}

The first two LHC runs have firmly established the Standard Model (SM) as the correct description of electroweak interactions at the energies explored so far. The discovery of a Higgs boson~\cite{higgs}, with couplings consistent with SM expectations, has completed the SM spectrum, and no additional states have been observed. This absence of new physics hints at a mass gap between the SM and any possible heavy degrees of freedom, motivating the use of effective field theories (EFTs) to probe indirect signals of higher scales at low energies.
Constructing an EFT requires specifying the particle content, symmetries, and power counting~\cite{EFT}. In the electroweak sector, the choice of power counting depends on how the Higgs field is introduced. One can adopt the linear realization~\cite{Brivio:2017vri}, where the Higgs $h$ belongs to an $SU(2)$ doublet with the three Goldstone bosons, or the more general non-linear realization~\cite{EFT}, where no specific relation between the Higgs and the Goldstones $\vec{\varphi}$ is assumed. In this work, we follow the non-linear approach~\cite{lagrangian}, which encompasses the linear case as a particular limit.

Beyond the SM particle content, we consider a strongly coupled scenario with heavy bosonic resonances ($J^P=0^\pm,1^\pm$) interacting with SM fields. In previous studies we have analyzed the contributions of these resonances to the low-energy constants (LECs) of the electroweak effective Lagrangian~\cite{PRD}. Here, we focus on the constraints that the electroweak oblique parameters $S$ and $T$~\cite{Peskin_Takeuchi} impose on the resonance spectrum. The parameter $S$ encodes the new-physics (NP) contribution to the difference between the $Z$-boson self-energy evaluated at $Q^2 = M_Z^2$ and at $Q^2 = 0$. In contrast, $T$ is proportional to the NP-induced difference between the $W$- and $Z$-boson self-energies at $Q^2 = 0$. By construction, both parameters vanish within the SM, while in the presence of NP they are expected to be of $\mathcal{O}(1)$.
In an earlier work, we performed a one-loop calculation of $S$ and $T$ in this framework~\cite{ST}, extending our previous Higgsless analysis~\cite{S_Higgsless} by including the Higgs boson. That study showed that precision electroweak data require the $hWW$ coupling to be extremely close to the SM prediction, while vector and axial-vector resonances must be nearly degenerate and heavier than about $4\,$TeV. At the time, these bounds were stronger than direct LHC limits.
Thanks to the much larger data sets now available, the $hWW$ coupling has been determined with high precision, $\kappa_W=1.023\pm 0.026$ (in SM units)~\cite{PDG}. Here we use the measured value as input rather than treating $\kappa_W$ as a free parameter~\cite{new}. This allows us to refine the previous analysis and extend it to a broader set of interactions~\cite{new}, including both $P$-even and $P$-odd operators, beyond the P-even sector considered earlier.

The oblique parameters can be expressed through convergent dispersive relations involving spectral functions~\cite{Peskin_Takeuchi,ST}. At leading order, they are saturated by vector and axial-vector exchange ($T=0$ at this level). Next-to-leading order corrections are dominated by the lightest two-particle cuts, while multi-particle effects with heavy resonances are negligible. In addition to the Goldstone–Goldstone  ($\varphi\varphi$) and Higgs–Goldstone ($h\varphi$) contributions studied before, we now also include corrections from light fermion–antifermion intermediate states ($\psi \bar{\psi}$)~\cite{new}.

\section{The calculation}

The effective Lagrangian is not organized by the canonical dimensions of the operators; instead, they are classified according to the chiral dimension $\hat d$, which characterizes the infrared behavior at low momenta~\cite{Weinberg}. Since our focus is on the next-to-leading-order (NLO) resonance contributions to the $S$ and $T$ parameters arising solely from SM cuts, it suffices to restrict the analysis to $\mathcal{O}(p^2)$ operators containing at most one spin-$1$ bosonic resonance field. Using the notation of Ref.~\cite{lagrangian}, the relevant Lagrangian for the calculations reads as~\cite{new}
\begin{align}
\Delta \mathcal{L}_{\mathrm{RT}} &=\,\quad 
\sum_{\xi} \left[ i\,\bar\xi \gamma^\mu d_\mu \xi  
 - v \left(\bar{\xi}_L \mY \xi_R + \mbox{h.c.}\right)  
\right]  \,+\, \frac{v^2}{4}\,\left( 1 +\Frac{2\kappa_W}{v} h \right) \bra u_\mu u^\mu\ket_{2} \nonumber \\
&+\,\bra V^1_{3\,\mu\nu} \left( \Frac{F_V}{2\sqrt{2}}  f_+^{\mu\nu} + \Frac{i G_V}{2\sqrt{2}} [u^\mu, u^\nu]  + \Frac{\widetilde{F}_V }{2\sqrt{2}} f_-^{\mu\nu}  +  \Frac{ \widetilde{\lambda}_1^{hV} }{\sqrt{2}}\left[  (\partial^\mu h) u^\nu-(\partial^\nu h) u^\mu \right]   + C_{0}^{V^1_3} J_T^{\mu\nu}  \right) \ket_2 \nonumber \\
&+ \,\bra A^1_{3\,\mu\nu} \left(\Frac{F_A}{2\sqrt{2}}  f_-^{\mu\nu}  + \Frac{ \lambda_1^{hA} }{\sqrt{2}} \left[ (\partial^\mu h) u^\nu-(\partial^\nu h) u^\mu \right] +  \Frac{\widetilde{F}_A}{2\sqrt{2}} f_+^{\mu\nu} +  \Frac{i \widetilde{G}_A}{2\sqrt{2}} [u^{\mu}, u^{\nu}]   +  \widetilde{C}_{0}^{A^1_3}  J_{T}^{\mu\nu} \right) \ket_2 \, ,
%
%
\label{eq:Lagr}
\end{align} 
where $V^1_{3\,\mu\nu}$ and $A^1_{3\,\mu\nu}$ denote color-singlet, custodial-triplet resonances with quantum numbers $J^{PC}=1^{--}$ ($V$) and $1^{++}$ ($A$), respectively, described within the antisymmetric formalism~\cite{RChT}. Note that $\bra ... \ket_2$ indicates an $SU(2)$ trace. The Goldstone fields are encoded in the $SU(2)$ matrix $U=u^2=\exp{(i\,\vec{\sigma}\cdot\vec{\varphi}/v)}$, with $v=(\sqrt{2}G_F)^{-1/2}=246\;\mathrm{GeV}$ the electroweak symmetry breaking scale. The building block $u_\mu=-i\,u^\dagger D_\mu U\,u^\dagger$ involves the covariant derivative $D_\mu$, while $f_\pm^{\mu\nu}$ contains the gauge field strengths. Couplings with a tilde correspond to odd-parity operators, which are assumed in this work to be suppressed relative to their even-parity counterparts.

The effective resonance Lagrangian in (\ref{eq:Lagr}) involves ten resonance couplings plus two resonance masses. To reduce the number of free resonance parameters—and thereby obtain meaningful phenomenological bounds—we impose high-energy constraints. Since this Lagrangian interpolates between the low- and high-energy regimes, ensuring a consistent short-distance behavior is also theoretically essential. The following constraints have been implemented:
\begin{itemize}
\item {\bf Form factors}. Requiring the two-Goldstone ($\varphi\varphi$) and Higgs–Goldstone ($h\varphi$) vector and axial form factors to vanish at high energies fixes the parameters $G_V$, $\widetilde{G}_A$, $\lambda_1^{hA}$, and $\widetilde{\lambda}_1^{hV}$ in terms of the remaining ones~\cite{PRD,new},
\begin{align}
\frac{G_V}{F_A} = - \frac{\widetilde G_A}{\widetilde{F}_V} = \frac{\lambda_1^{hA} v }{\kappa_W F_V} = -\frac{\widetilde{\lambda}_1^{hV} v}{\kappa_W \widetilde{F}_A} = \frac{v^2}{F_V F_A - \widetilde{F}_V \widetilde{F}_A}\, . \label{constraints_summary}
\end{align}
\item {\bf Weinberg sum rules (WSRs)}. The assumed chiral symmetry of the underlying electroweak theory implies that the $W^3 B$ correlator acts as an order parameter of electroweak symmetry breaking, vanishing at high energies in asymptotically free gauge theories as $1/s^3$. This leads to the first and second Weinberg sum rules~\cite{WSR}, corresponding to the vanishing of the $1/s$ and $1/s^2$ terms, respectively. In this case, this translates into the following relations~\cite{new},
\begin{equation}
 \left( F_V^{2} \!-\! \widetilde{F}_V^{2} \right)  \!-\! \left( F_A^{2}\!-\! \widetilde{F}_A^{2}\right)   = v^2  \left( 1\!+\! \delta_{_{\rm NLO}}^{(1)} \right) \, , \qquad %
 \left( F_V^{2} \!-\! \widetilde{F}_V^{2} \right) M_V^{2} \!-\! \left( F_A^{2}\!-\! \widetilde{F}_A^{2}\right) M_A^{2} = v^2 \,M_V^{2}\,  \delta_{_{\rm NLO}}^{(2)}  \,, \label{WSR}
\end{equation}
where $ \delta_{_{\rm NLO}}^{(1)}$ and $ \delta_{_{\rm NLO}}^{(2)}$ encode the one-loop contributions. Additionally, the 2nd WSR implies the vanishing of $\widetilde\delta_{_{\rm NLO}}^{(1)}$ and $\widetilde\delta_{_{\rm NLO}}^{(2)}$~\cite{new}, but there is no contribution to $\widetilde\delta_{_{\rm NLO}}^{(1)}$ in this case.  While the 1st WSR, left constraint in (\ref{WSR}), is generally expected to hold in gauge theories with nontrivial ultraviolet (UV) fixed points, the validity of the 2nd WSR, right expression in (\ref{WSR}), depends on the specific UV dynamics. Imposing both WSRs determines the combinations $F_V^2 - \widetilde{F}_V^2$ and $F_A^2 - \widetilde{F}_A^2$, which play a central role in fixing the $S$ and $T$ parameters. If only the 1st WSR is assumed, one can still determine $T$ and derive a lower bound on $S$~\cite{new}.

It should be emphasized that, in the absence of $P$-odd couplings, the combined implementation of the 1st and 2nd WSRs implies the mass hierarchy $M_A > M_V$. This relation continues to hold under the reasonable assumption that odd-parity couplings are suppressed relative to their even-parity counterparts ($|\widetilde{F}_{V,A}|\!<\!|F_{V,A}|$), which we adopt here and consequently an expansion in terms of $\widetilde{F}_V/F_V$ and $\widetilde{F}_A/F_A$ is performed. Accordingly, we will assume $M_A\! >\! M_V$ in all dynamical scenarios, even in cases where the 2nd WSR is not applicable.

Moreover, considering again that odd-parity couplings are smaller than the corresponding even-parity ones, the 1st WSR together with the upper bound on the vector coupling $(C_0^{V^1_3})^2$, extracted from LHC diboson production studies ($WW$, $WZ$, $ZZ$, $Wh$, and $Zh$)~\cite{lagrangian,Dorigo:2018cbl}, allows us to conclude that fermionic contributions are negligible.
\end{itemize}
The use of (\ref{constraints_summary}) and (\ref{WSR}), the neglect of the fermionic contributions, and the expansion in terms of $\widetilde{F}_{V,A}/F_{V,A}$ allows us to determine $S$ and $T$ in terms of only the resonance masses ($M_V$ and $M_A$) and the expansion parameters $\widetilde{F}_{V,A}/F_{V,A}$, so that the phenomenology turns into a promising task. We have considered two scenarios: underlying theories satisfying the two WSRs and the more conservative case where only the 1st WSR is assumed. In the first case, $S$ is given by~\cite{new}
\begin{eqnarray}
S_{\mathrm{NLO}}  &=&  4\pi v^2  \bigg(\!\frac{1}{M_{V}^{2}} +\frac{1}{M_{A}^{2}}\bigg) + \Delta S_{\mathrm{NLO}}^{\mathrm{P-even}} +  \Delta S_{\mathrm{NLO}}^{\mathrm{P-odd}} \,  , \nonumber \\
 \Delta S_{\mathrm{NLO}}^{\mathrm{P-even}} &=& \frac{1}{12\pi} \left[ \left(1-\kappa_W^2\right)  \left( \log \frac{M_V^2}{m_h^2} -\frac{11}{6} \right)  +\kappa_W^2 \left( \frac{M_A^2}{M_V^2}-1 \right) \log \frac{M_A^2}{M_V^2}  \right]\,, \nonumber  \\
 \Delta S_{\mathrm{NLO}}^{\mathrm{P-odd}}  &=&
\frac{1}{12\pi}   \left(\frac{\widetilde{F}_V^2}{F_V^2} +2\kappa_W^2 \frac{\widetilde{F}_V \widetilde{F}_A}{F_V F_A} - \kappa_W^2 \frac{\widetilde{F}_A^2}{F_A^2} \right) \left( \frac{M_A^2}{M_V^2} - 1\right)  \log \frac{M_A^2}{M_V^2} 
  + \mathcal{O}\!\left(\!\frac{\widetilde{F}^4_{V,A}}{F^4_{V,A}}\!\right) 
\, ,  \label{SNLO}
\end{eqnarray}
where, from now on, we split the result in three contributions: tree-level, one-loop including only $P$-even operators and one-loop including $P$-odd operators. $T$ reads as~\cite{new}
\begin{eqnarray}
T_{\mathrm{NLO}}  &=&  \Delta T_{\mathrm{NLO}}^{\mathrm{P-even}} +  \Delta T_{\mathrm{NLO}}^{\mathrm{P-odd}} \,  , \nonumber \\
\Delta T_{\mathrm{NLO}}^{\mathrm{P-even}}  &=& \frac{3}{16\pi \cos^2 \theta_W} \left[ \left( 1-\kappa_W^2 \right) \left( 1 - \log{\frac{M_V^2}{m_{h}^2}} \right)  + \kappa_W^2 \log{\frac{M_A^2}{M_V^2}}   \right]   , \nonumber \\
\Delta T_{\mathrm{NLO}}^{\mathrm{P-odd}} &=&   \frac{3}{16\pi \cos^2 \theta_W} 
\Bigg\{  2\kappa_W^2  \frac{\widetilde{F}_A}{F_A}- 2\frac{\widetilde{F}_V}{F_V} +  \frac{M_V^2}{M_A^2-M_V^2} 
 \log \frac{M_A^2}{M_V^2} \left( 2\frac{\widetilde{F}_V}{F_V} - 2 \kappa_W^2 \frac{M_A^2}{M_V^2} \frac{\widetilde{F}_A}{F_A}\right)    \nonumber \\ && 
 +  \frac{M_V^2}{M_A^2-M_V^2} 
 \log \frac{M_A^2}{M_V^2} \left[  
 \left(\kappa_W^2 \frac{\widetilde{F}_A^2}{F_A^2} - \frac{\widetilde{F}_V^2}{F_V^2}  \right)
 \left(1+ \frac{M_A^2}{M_V^2} \right) 
 +2 \frac{ \widetilde{F}_V\widetilde{F}_A}{ F_VF_A } \left( \kappa_W^2 \frac{M_A^2}{M_V^2} - 1 \right) 
  \right]\nonumber \\ &&
    +2 \left(
    \frac{\widetilde{F}_V^2}{F_V^2} -
    \kappa_W^2 \frac{\widetilde{F}_A^2}{F_A^2}  + \left( 1 - \kappa_W^2 \right)\frac{ \widetilde{F}_V\widetilde{F}_A}{ F_VF_A } \right) \Bigg\}
 + \mathcal{O}\left(\frac{\widetilde{F}^3_{V,A}}{F^3_{V,A}}\right)\,
.\label{TNLO}  
\end{eqnarray}
Finally, in this scenario assuming both WSRs one has the one-loop constraint $\widetilde\delta_{_{\rm NLO}}^{(2)}=0$, being~\cite{new}
\begin{eqnarray}
\widetilde\delta_{_{\rm NLO}}^{(2)}\!&\!=\!&\! \widetilde\delta_{_{\rm NLO}}^{(2)}\big|^{\mathrm{P-even}} + \widetilde\delta_{_{\rm NLO}}^{(2)}\big|^{\mathrm{P-odd}} \,, \nonumber \\
\widetilde\delta_{_{\rm NLO}}^{(2)}\big|^{\mathrm{P-even}}\!&\!=\!&\!  \frac{M_V^2}{48\pi^2v^2} \left( -1 +\kappa_W^2 \frac{M_A^4}{M_V^4}  \right) \,, \nonumber \\
\widetilde\delta_{_{\rm NLO}}^{(2)}\big|^{\mathrm{P-odd}}\!&\!=\!&\!  
\frac{M_A^2 - M_V^2}{48\pi^2v^2} \left[ \left( \frac{M_A^2}{M_V^2}-1 \right)  
\left( \frac{\widetilde{F}_V^2}{F_V^2}-\kappa_W^2  \frac{\widetilde{F}_A^2}{F_A^2} \right) +  \frac{2\widetilde{F}_V \widetilde{F}_A}{F_V F_A} \left( 1 + \kappa_W^2 \frac{M_A^2}{M_V^2} \right) \right] + \mathcal{O}\left(\!\frac{\widetilde{F}^4_{V,A}}{F^4_{V,A}}\!\right) .
 \label{tildedelta2}
\end{eqnarray}
If the 2nd WSR is not satisfied, $T$ does not change and a lower bound on $S$ is obtained~\cite{new},
\begin{eqnarray}
  S_{\mathrm{NLO}} \!\!&\!\!>\!\!&\!\!  \frac{4\pi v^2}{M_V^{2}} 
+ \Delta \widetilde{S}_{\mathrm{NLO}}^{\,\mathrm{P-even}} +  \Delta \widetilde{S}_{\mathrm{NLO}}^{\,\mathrm{P-odd}} \,  , \nonumber \\
 \Delta \widetilde{S}_{\mathrm{NLO}}^{\,\mathrm{P-even}} \!\!&\!\!=\!\!&\!\! 
  \frac{1}{12\pi}
\left[     \bigg(1-\kappa_W^2 \bigg) \bigg(\log\frac{M_V^2}{m_{h}^2}-\frac{11}{6}\bigg) 
 - \,\kappa_W^2\, \bigg(\log \frac{M_A^2}{M_{V}^2}-1
 + \frac{M_A^2}{M_V^2}\bigg) \right]  \,, \nonumber \\
 \Delta \widetilde{S}_{\mathrm{NLO}}^{\,\mathrm{P-odd}}   \!\!&\!\!=\!\!&\!\! 
    \frac{1}{12\pi} \left\{\! \left(\!1\!-\!\frac{M_A^2}{M_V^2}\!\right) \!    \left[\! \frac{\widetilde{F}_V^2}{F_V^2}\!+\! \kappa_W^2 \frac{\widetilde{F}_A}{F_A}\left( \! 2 \frac{\widetilde{F}_V}{F_V}\!-\!\frac{\widetilde{F}_A}{F_A} \!\right)\!\right]   \!+\! \log \frac{M_A^2}{M_V^2}\!\left( \!\frac{\widetilde{F}_V^2}{F_V^2}\!-\! \kappa_W^2 \frac{\widetilde{F}_A^2}{F_A^2} \!-\! 2 \frac{\widetilde{F}_V\widetilde{F}_A}{F_VF_A}\!\right)\!
    \right\} 
    \!+\! \mathcal{O}\!\left(\!\frac{\widetilde{F}^4_{V,A}}{F^4_{V,A}}\!\right) . \nonumber \\ && \label{SNLOB}
\end{eqnarray}

\section{Phenomenology}

\begin{figure*}
\begin{center}
\includegraphics[scale=0.21]{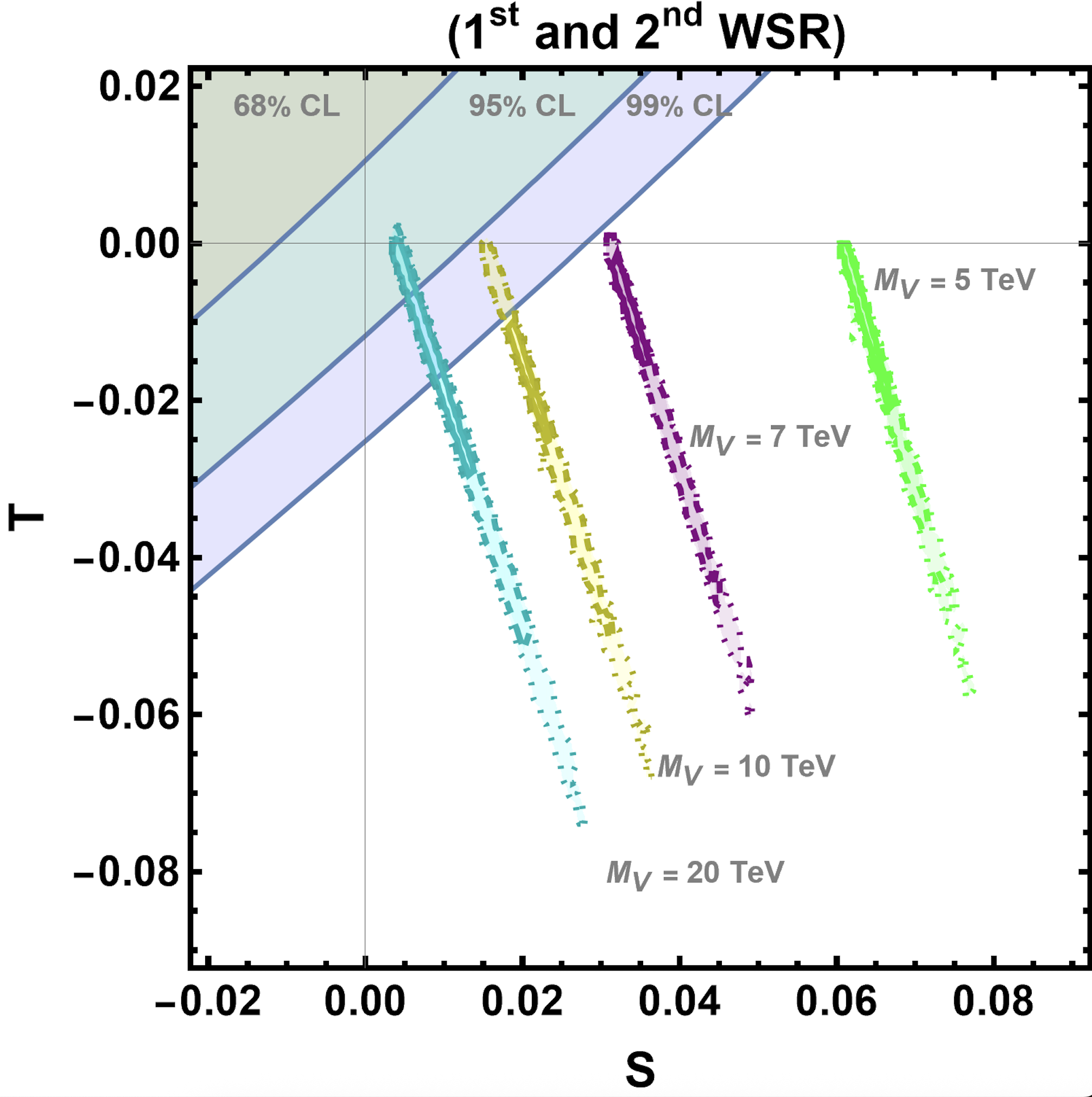} 
\includegraphics[scale=0.21]{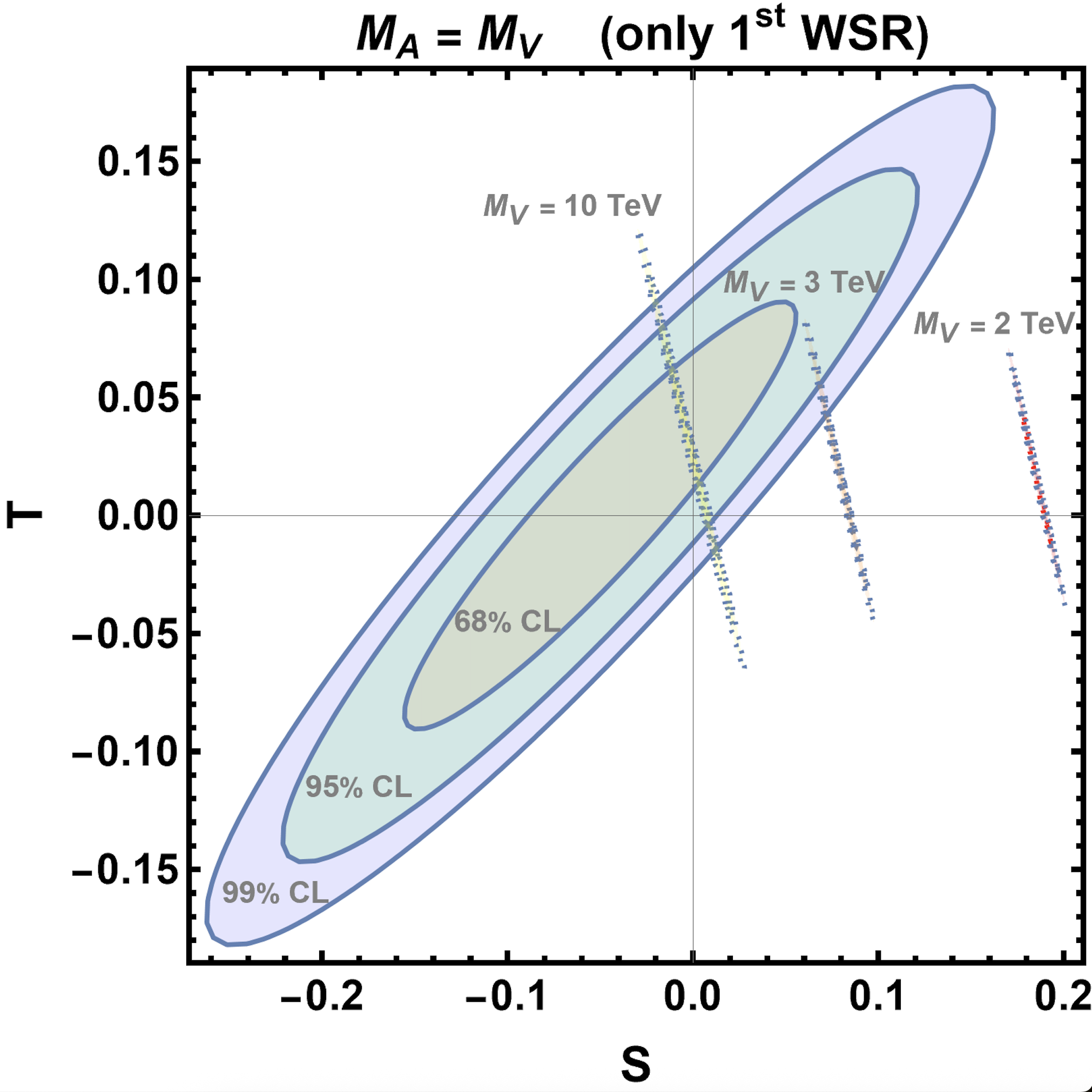} 
\includegraphics[scale=0.21]{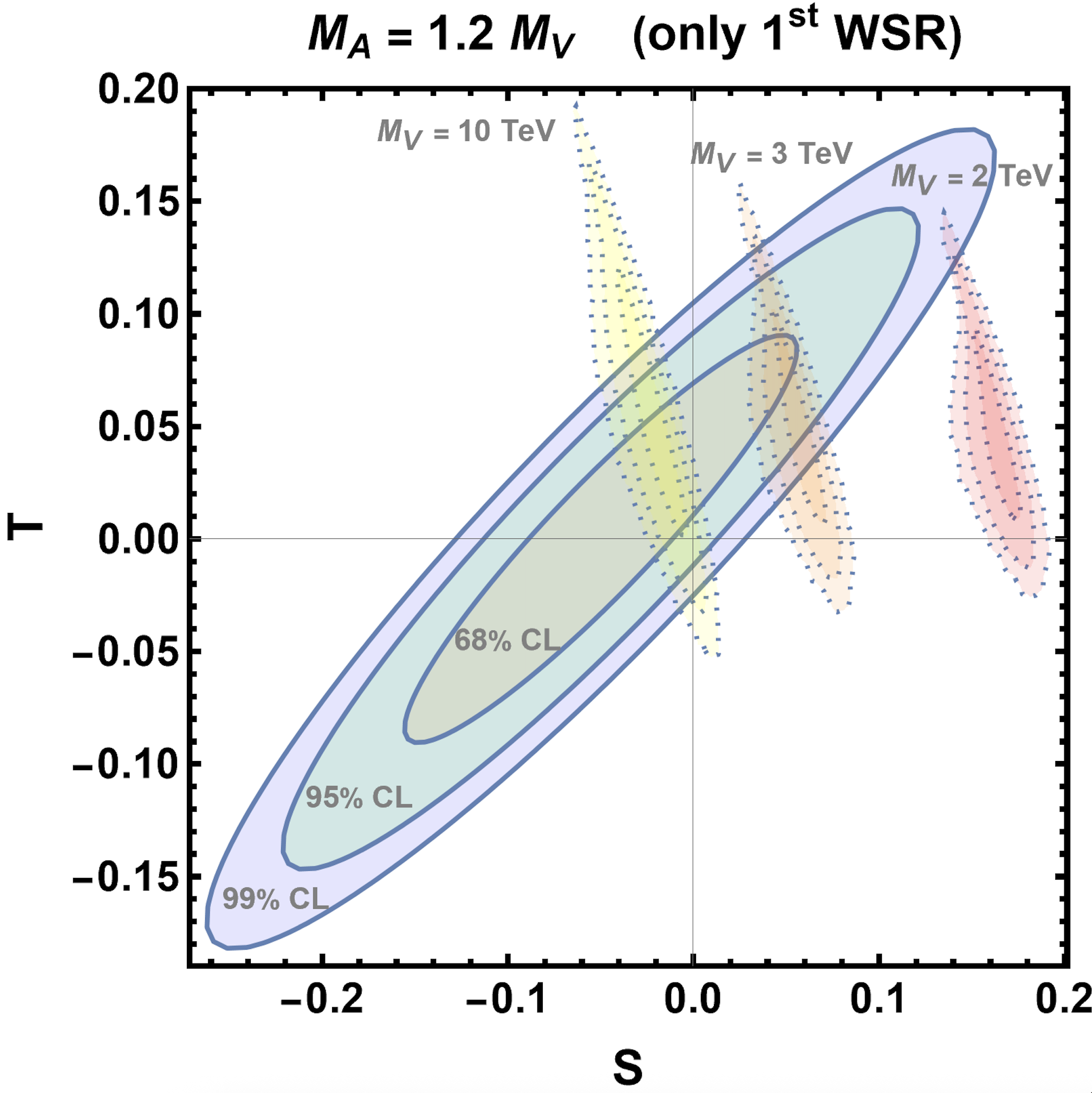} 
\includegraphics[scale=0.21]{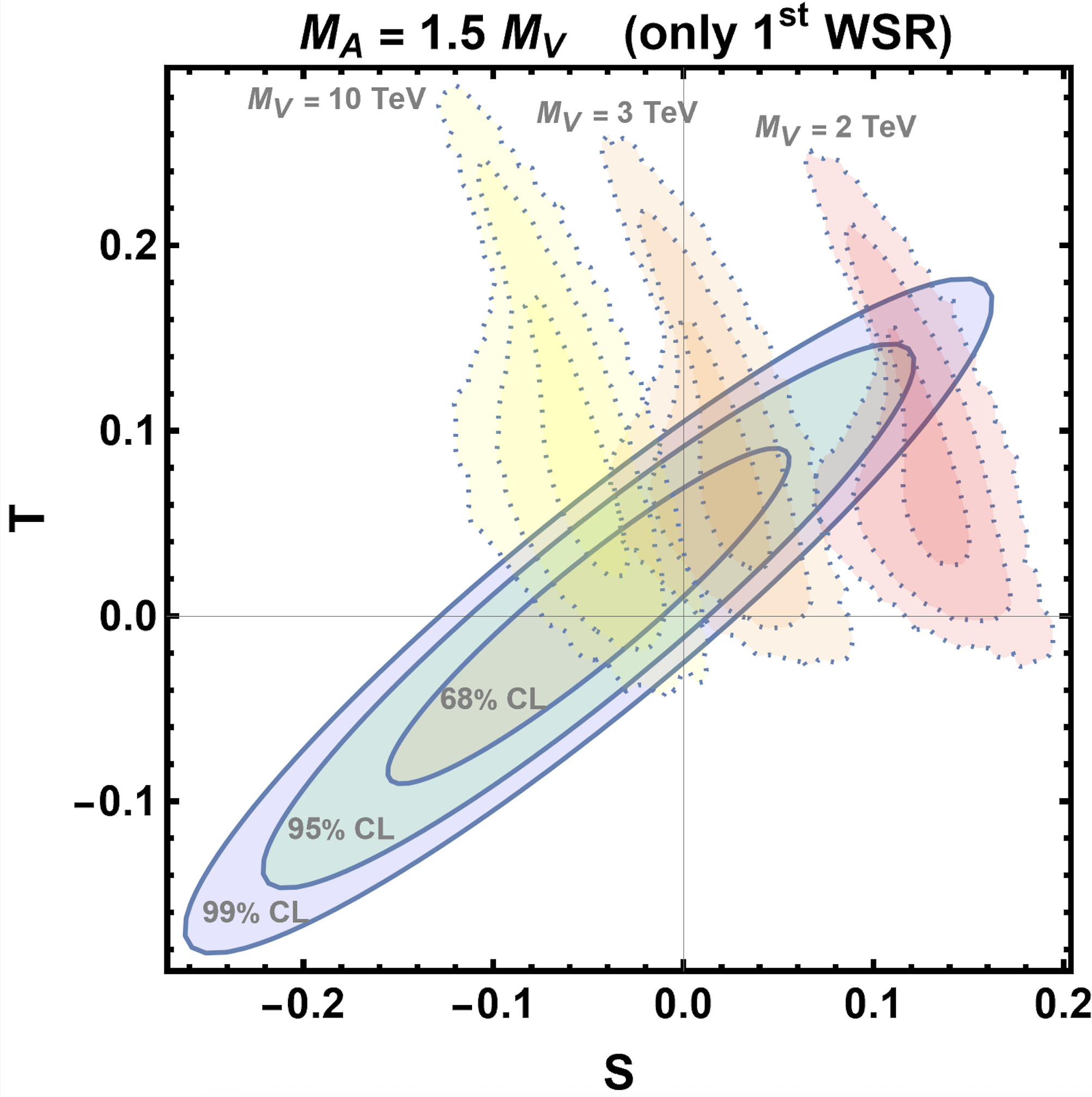}
\caption{{\small
Next-to-leading-order predictions of $S$ and $T$ considering both WSRs (top-left plot) or only the 1st WSR (top-right and bottom plots), so only lower bounds on $S$ are shown in the second case. The ellipses give the experimentally allowed regions of $S$ and $T$ at $68$\%, $95$\% and $99$\% CL~\cite{PDG}. The different colors of the points correspond to different values of $M_V$: $M_V=2$ (red), $3$ (orange), $5$ (green), $7$ (purple), $10$ (yellow) and $20$ (blue) TeV. For the different cases, the predictions are plotted at $68$\%, $95$\% and $99$\% CL. When the 2nd WSR is assumed, $M_A$ is determined by $\widetilde\delta_{_{\rm NLO}}^{(2)}=0$, whereas we have considered different values for $M_A$ in terms of $M_V$ when the 2nd WSR is discarded: $M_A=M_V$ (top-right), $M_A=1.2\; M_V$ (bottom-left) and $M_A=1.5\; M_V$ (bottom-right). The values of $\kappa_W$ and $\widetilde{F}_{V,A}/F_{V,A}$   have been obtained by considering normal distributions given by $\kappa_W=1.023\pm 0.026$~\cite{PDG} and $\widetilde{F}_{V,A}/F_{V,A}=0.00\pm 0.33$. }}
\label{fig:NLO}
\end{center}
\end{figure*}

In this section, we derive constraints on the masses of possible NP resonances by confronting our theoretical predictions of (\ref{SNLO})-(\ref{SNLOB}) with the experimental determinations of the oblique parameters $S$ and $T$. We take the values from the PDG~\cite{PDG}: $S=-0.05\pm 0.07$ and $T=0.00 \pm 0.06$, with a correlation coefficient of 0.93, together with $\kappa_W=1.023\pm 0.026$.

As it has been explained previously, for this one-loop analysis, several approximations have been made: only the lightest absorptive cuts are retained ($\varphi\varphi$, $h\varphi$, and $\psi \bar{\psi}$); odd-parity couplings are treated as subleading, enabling an expansion in $\widetilde{F}_{V,A}/F_{V,A}$; and the mass ordering $M_A > M_V$ is assumed. In fact, when both WSRs are imposed, the latter is not merely an assumption but a theoretical requirement.

Note that the theoretical determinations of (\ref{SNLO})-(\ref{SNLOB}) are given in terms of only four free resonance parameters: $M_V$, $M_A$, $\widetilde{F}_V/F_V$ and $\widetilde{F}_A/F_A$. Taking into account that the last two resonance parameters are supposed to be small, we consider a normal distribution $\widetilde{F}_{V,A}/F_{V,A}=0.00\pm 0.33$. Moreover, in the case of assuming both WSRs, the constraint $\widetilde\delta_{_{\rm NLO}}^{(2)}=0$ allows us to determine $M_A$ in terms of $M_V$, giving rise to a value close to $M_V$. 

The comparison between our NLO determinations of $S$ and $T$ and the experimental values are shown in Figure~\ref{fig:NLO}~\cite{new}. The ellipses give the experimentally allowed regions of $S$ and $T$ at $68$\%, $95$\% and $99$\% CL~\cite{PDG}. The different values of $M_V$ are depicted with different colors: $M_V=2$ (red), $3$ (orange), $5$ (green), $7$ (purple), $10$ (yellow) and $20$ (blue) TeV; being the regions plotted at $68$\%, $95$\% and $99$\% CL. The top-left plot corresponds to the case assuming both WSRs and it implies~\cite{new}
\begin{equation}
M_A\, \ge \, M_V \gsim  10\;\mathrm{TeV} \quad (95\%\; \mathrm{CL}). \label{masses_2WSR}
\end{equation}
If the underlying theory does not satisfy the 2nd WSR, $M_A$ is a free parameter which is not expected to be very far from $M_V$ and we show the cases $M_A=M_V$ (top-right), $M_A=1.2\,M_V$ (bottom-left) and $M_A=1.5\,M_V$ (bottom-right). These results indicate that, in scenarios where the 2nd WSR does not hold, current electroweak precision data remain compatible with lower values for the heavy resonances~\cite{new}, 
\begin{equation}
M_{A}\, \geq \, M_V \, \gsim  2 \;\mathrm{TeV} \quad (95\%\; \mathrm{CL}). \label{masses_1WSR}
\end{equation}

In summary, we find that the $P$-odd operators and the fermionic-cut contributions considered here induce only small corrections to the oblique parameters and therefore to our previous bounds on $M_{V,A}$ in scenarios restricted to $P$-even operators. These results reinforce and extend the conclusions of earlier studies~\cite{ST}, providing further support for their validity.


\end{document}